\documentclass[aps,prl,twocolumn,superscriptaddress,showpacs]{revtex4-1}

\usepackage{color}
\usepackage{graphicx}
\usepackage{array}
\usepackage{gensymb}

\begin{document}
\title{Reversible \textit{vs.} irreversible voltage manipulation of interfacial magnetic anisotropy in Pt/Co/oxide multilayers  }
\
\author{Aymen Fassatoui}

\affiliation{Univ.~Grenoble Alpes, CNRS, Institut N\'{e}el, Grenoble, France}
\
\author{Jose Pe\~{n}a Garcia}
\affiliation{Univ.~Grenoble Alpes, CNRS, Institut N\'{e}el, Grenoble, France}

\author{Jan Vogel}
\affiliation{Univ.~Grenoble Alpes, CNRS, Institut N\'{e}el, Grenoble, France}
\
\author{Laurent Ranno}
\affiliation{Univ.~Grenoble Alpes, CNRS, Institut N\'{e}el, Grenoble, France}

\author{Anne Bernand-Mantel}
\affiliation{Universit\'{e} de Toulouse, Laboratoire de Physique et Chimie des Nano-Objets, Toulouse, France}
\

\author{H\'{e}l\`{e}ne B\'{e}a}
\affiliation{Univ.~Grenoble Alpes, CEA, CNRS, Grenoble INP, IRIG-SPINTEC, Grenoble, France}
\
\author{Sergio Pizzini}
\affiliation{Department of Materials Science, University of Milano-Bicocca,  Via Cozzi 53, Milano, Italy}
\
\author{Stefania~Pizzini} \email{stefania.pizzini@neel.cnrs.fr}
\affiliation{Univ.~Grenoble Alpes, CNRS, Institut N\'{e}el, Grenoble, France}


\date{\today}

\begin{abstract}
The perpendicular magnetic anisotropy at the Co/oxide interface in Pt/Co/MO$_{x}$ (MO$_{x}$ = MgO$_{x}$, AlO$_{x}$, TbO$_{x}$) was modified by an electric field using a 10~nm-thick ZrO$_{2}$ as a solid electrolyte. The large voltage-driven modification of interfacial magnetic anisotropy  and the non-volatility of the effect is explained in terms of  the migration of oxygen ions towards/away from the Co/MO$_{x}$ interface. While the effect is reversible in Pt/Co/AlO$_{x}$ and Pt/Co/TbO$_{x}$, where the Co layer can be oxidised or reduced, in Pt/Co/MgO$_{x}$ the effect has been found to be irreversible.  We  propose that these differences may be related to the different nature of the ionic conduction within the MO$_{x}$ layers.

\end{abstract}

\maketitle

\section{Introduction}
The manipulation of the interfacial magnetic anisotropy \textit{via} an electric-field is  an active field of research, as it is a promising route towards the realisation of low power spintronic devices \cite{Nozaki2019}. In metallic ferromagnetic (FM) thin films, electric fields have been shown to modify substantially the interface magnetic anisotropy energy through the modification of the electron density of states at the Fermi energy, despite their short penetration depth limited by Coulomb screening \cite{Weisheit2007,Maruyama2009,Shiota2011,Matsukura2015}.
The perpendicular  magnetic anisotropy (PMA)  at a FM/oxide interface  can be as large as the one found at a heavy metal/Co interface  \cite{Monso2002}. \textit{Ab initio} calculations  suggest that this anisotropy results from the hybridisation of the oxygen and transition metal electronic orbitals across the interface \cite{Nozaki2019}. Several experimental studies of Pt/Co/MO$_{x}$ trilayers (M = Al, Mg, Ta, etc)  showed that the PMA amplitude is related to the degree of oxidation of the Co layer. The interfacial anisotropy constant (K$_{s}^{ox}$) at the Co/MO$_{x}$ interface  has a characteristic bell-like shape, with a maximum for the optimal oxidation conditions (see sketch in Fig.~\ref{fig:Figure1}), obtained when the oxygen atoms reach the Co-M interface, so that Co-O bonds prevail over Co-M bonds \cite{Manchon2008a,Manchon2008b,Manchon2008c,Dieny2017}. It is then not surprising that the
largest effects of electric-field on the PMA ($\Delta$K$_{s}$/E=$\beta>$1000~fJ/Vm) have been obtained by triggering the migration of oxygen ions towards/away from the FM/oxide interface \cite{Bi2014,Bauer2015,Zhou2016}. This magneto-ionic effect, obtained in most cases using Gd$_{2}$O$_{3}$ as dielectric oxide layer, leads to a non-volatile modification of the interfacial PMA, as opposed to the volatile effect  obtained by electron accumulation/depletion. Also, the timescales associated with electric field induced displacement of electrons or ions are different.

In the present work, we describe the effect of an electric field on the interfacial PMA in a series of Pt/Co/MO$_{x}$ (M=Mg, Al, Tb) trilayers. Beyond their interest because of their tunable PMA, these non-centrosymmetric systems have been largely studied  because they can host chiral magnetic textures such as chiral domain walls (DWs) and magnetic skyrmions, due to the presence of interfacial Dzyaloshinskii-Moriya  interaction (DMI).  It has been shown that, by tuning the PMA \cite{Bauer2012,Bauer2013,Chiba2012,Schellekens2012,Lin2016, Bernand-Mantel2013,Schott2016} and/or the DMI at the FM/oxide interface, \cite{Srivastava2018, HerreraDiez2019,Koyama2018} the electric fields can modify the stability and the field- and current- driven dynamics of such magnetic textures.

Using a 10~nm thick ZrO$_{2}$ as dielectric layer, we demonstrate that the effect of the electric field on the PMA is large ($\beta>$1200~fJ/Vm at room temperature) and non-volatile. Tuning the magnetic anisotropy allowed us to stabilise a variety of magnetic configurations within the Co layers and to modify the details of the magnetic reversal mechanisms. The non-volatility of the magnetic configurations obtained after the removal of the gate voltage, the characteristic time evolution of the electric field effect and its large efficiency can be understood by considering the ZrO$_{2}$ as a solid electrolyte working as an oxygen ion conductor. While the effect is reversible in Pt/Co/AlO$_{x}$ and Pt/Co/TbO$_{x}$, where the Co layer can be oxidised or reduced, in Pt/Co/MgO$_{x}$ the effect has been found to be irreversible.  Once the PMA has increased by increasing the Co oxidation, the reverse process does not occur by polarisation inversion. We  propose that this different behaviour may be related to the different nature of the ionic conduction within the MO$_{x}$ layer.

\section{Methods: film deposition and patterning}
Pt(3)/Co(1)/Mg(0.7), Pt(3)/Co(1)/Al(1) and Pt(3)/Co(1.1-1.2)/Tb(0.6) trilayers  (thicknesses in nm) were deposited  by magnetron sputtering on Si/SiO$_{2}$ wafers. The Mg and Tb layers were oxidised in oxygen atmosphere while Al layer was oxidised in an oxygen plasma at room temperature. The trilayers were patterned into 1-50~$\mu$m wide strips by electron beam lithography and ion beam milling. A 10~nm thick ZrO$_{2}$ dielectric layer was then deposited on top of the samples by atomic layer deposition (ALD). Finally, following a second lithography step, a 6~nm thick Pt layer was evaporated on top of the strips, leading to a local gate. In the three magnetic stacks the Co layer was under-oxidised relative to the optimal oxidation corresponding to the maximum PMA (see \cite{Manchon2008a,Manchon2008b,Manchon2008c,Dieny2017} and sketches in Figs.\ref{fig:Figure1}-\ref{fig:Figure3}). For more details of the fabrication see the Supplemental Material.
\par
The electric field is applied across the Pt electrodes, with the top one being grounded. The bias voltage is defined as  V$_{g}$ = V$_{top}$ - V$_{bottom}$, in order to keep the usual sign convention. Unless specified,  the bias voltage was applied with the sample being at room temperature.
The magnetic configuration of the Co layer below the electrode regions was observed with magnetic force microscopy (MFM) for Pt/Co/MgO$_{x}$ and Pt/Co/TbO$_{x}$  and by polar magneto-optical Kerr microscopy for Pt/Co/AlO$_{x}$. The hysteresis loops were measured by polar magneto-optical Kerr effect (MOKE). Note that the hysteresis loops and the magnetic images were measured always with V$_{g}$ = 0,  either in the initial state or after the application of the electric field, once V$_{g}$ had been removed and the new magnetic state was imprinted in the Co layer.

\section{Non volatile and irreversible manipulation of interfacial PMA in P\lowercase{t}/C\lowercase{o}/M\lowercase{g}O\lowercase{x} }
\begin{figure}[ht]
\begin{center}
\includegraphics[width=9cm]{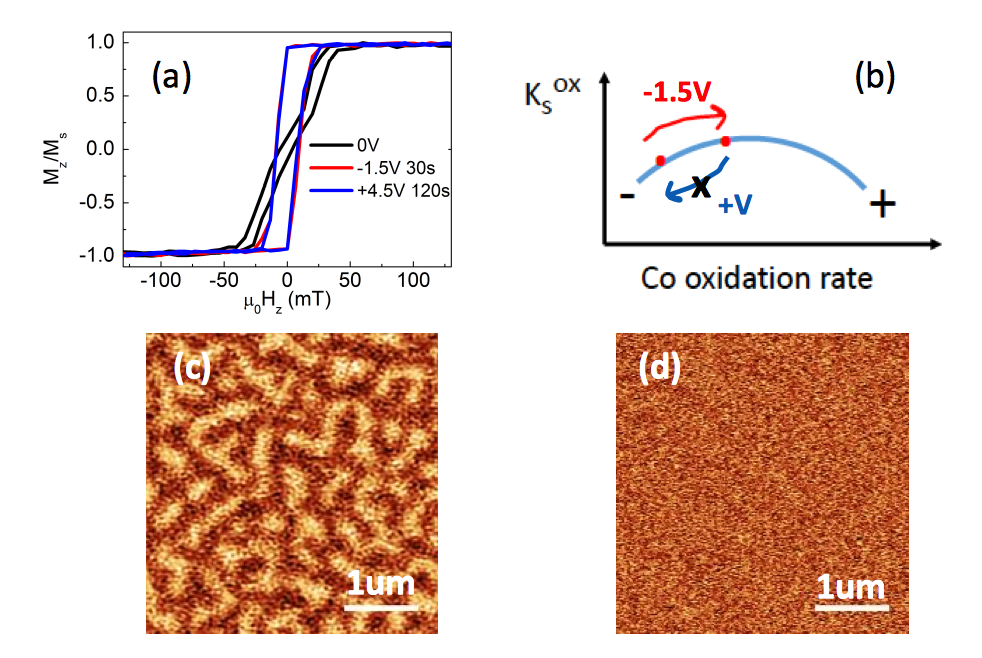}
\end{center}
\caption{\label{fig:Figure1} \textbf{Non-volatile and irreversible effect of the electric field in the Pt/Co/MgO$_{x}$ magnetic stack}
(a) Kerr hysteresis loops measured for Pt/Co/MgO$_{x}$ sample below the top Pt electrode,  in the initial state ($V_{g}$=0), after the application of  $V_{g}$=-1.5~V for 30~s, leading to an increase of the interfacial PMA, and after the application of  $V_{g}$=+4.5~V for 120~s, showing that the large PMA is maintained; (b) sketch representing the variation of the interfacial anisotropy constant K$_{s}^{ox}$ \textit{vs}. the Co layer oxidation rate \cite{Manchon2008a,Manchon2008b,Manchon2008c,Dieny2017} and the effect of the gate voltage; at the extremities of the K$_{s}^{ox}$  curve, (+) stands for over-oxidised Co, (-) for under-oxidised Co with respect to the optimal oxidation giving maximum PMA;  (c) MFM image showing the  labyrinthine domain configuration in the initial state; (d) MFM image after the application of  $V_{g}$=-1.5~V for 30~s. During the image scans, $V_{g}$=0 and no out-of-plane magnetic field is applied. Dark and light contrasts correspond to down and up magnetisation. }
\end{figure}

The hysteresis loop measured for the Pt/Co/MgO$_{x}$ trilayer exhibits a butterfly-like shape with low remanence, indicating   thermally activated formation of magnetic domains (Fig.~\ref{fig:Figure1}(a)).  The MFM image confirms the presence of labyrinthine domains with out-of-plane magnetisation, with an average width of $\simeq$120nm (Fig.~\ref{fig:Figure1}(c)).
\par
The formation of  labyrinthine domains is the consequence  of  the balance between demagnetising energy and domain wall energy in a perpendicularly magnetised system. This magnetic configuration can be observed  in the vicinity of the reorientation transition from in-plane (IP) to out-of-plane (OOP) magnetisation,  when the effective magnetic anisotropy $K_{eff}$ is close to zero and changes sign:
\begin{equation}
 K_{eff} = \frac{K_{s}}{t_{Co}} -\frac{1}{2}\mu_{0} M_{s}^{2} = \frac{K_{s}^{ox}+K_{s}^{Pt}}{t_{Co}} -\frac{1}{2}\mu_{0} M_{s}^{2} \simeq 0
 \end{equation}
 where $K_{s}$ is the interfacial anisotropy constant, with $K_{s}^{Pt}$ and $K_{s}^{ox}$ the contributions from the Co/Pt and Co/MO$_{x}$ interfaces and $M_{s}$ is the spontaneous magnetisation.

The labyrinthine domain width $L$ in the case of ultrathin ferromagnetic layers is given by \cite{Schafer1998,Kaplan1993}:
 \begin{equation}
 L=C~t~\exp\frac{\pi L_{0}}{t}
 \end{equation}
 where $L_{0}=\sigma/\mu_{0}M_{s}^{2}$ is the characteristic dipolar length; $\sigma$ is the domain wall energy, $t$ is the ferromagnetic film thickness and $C$ is a numerical constant of the order of 1. In the non-centrosymmetric stacks studied in this work, domain walls have chiral N\'{e}el structure \cite{Boulle2016,Juge2019} and their energy is given $\sigma=(4\sqrt{AK_{eff}}-\pi D)$,  where $A$ is the exchange stiffness and  $D$ is the strength of the Dzyaloshinskii-Moriya interaction.
The observed labyrinthine structure can be reproduced by micromagnetic simulations  using experimentally relevant magnetic parameters (see Supplemental Material).

After the application of  a bias voltage $V_{g}$=-1.5~V for 30 seconds,  the stripe domains disappear and a  magnetic state with 100\%  remanence at zero field is obtained, as shown by the square hysteresis loop  (Fig.~\ref{fig:Figure1}(a)) and the MFM image (Fig.~\ref{fig:Figure1}(d)). This indicates that the negative voltage drives the increase of the interfacial anisotropy anisotropy which, as shown in Fig.~\ref{fig:Figure1}(c)), can be associated to the increase of the Co oxidation.   The same magnetic state is maintained for several weeks after the removal of the electric field \textit{i.e.} the change of magnetic properties driven by the electric field can be defined as non-volatile.
Fig.~\ref{fig:Figure1}(a) shows that after the application of a reverse voltage $V_{g}$=+4.5~V for 2 minutes, the hysteresis loop stays unchanged.  In order to confirm the irreversibility of the electric-field effect, a much larger bias voltage ($V_{g}$=+12~V) was also applied  for several minutes and the sample temperature was increased  up to 200\degree C during the electric field application. For positive voltages, whatever the condition, the hysteresis loops remained square i.e.  the electric field effect appears to be irreversible.
\par

\begin{figure*}[ht]
\begin{center}
\includegraphics[width=18cm]{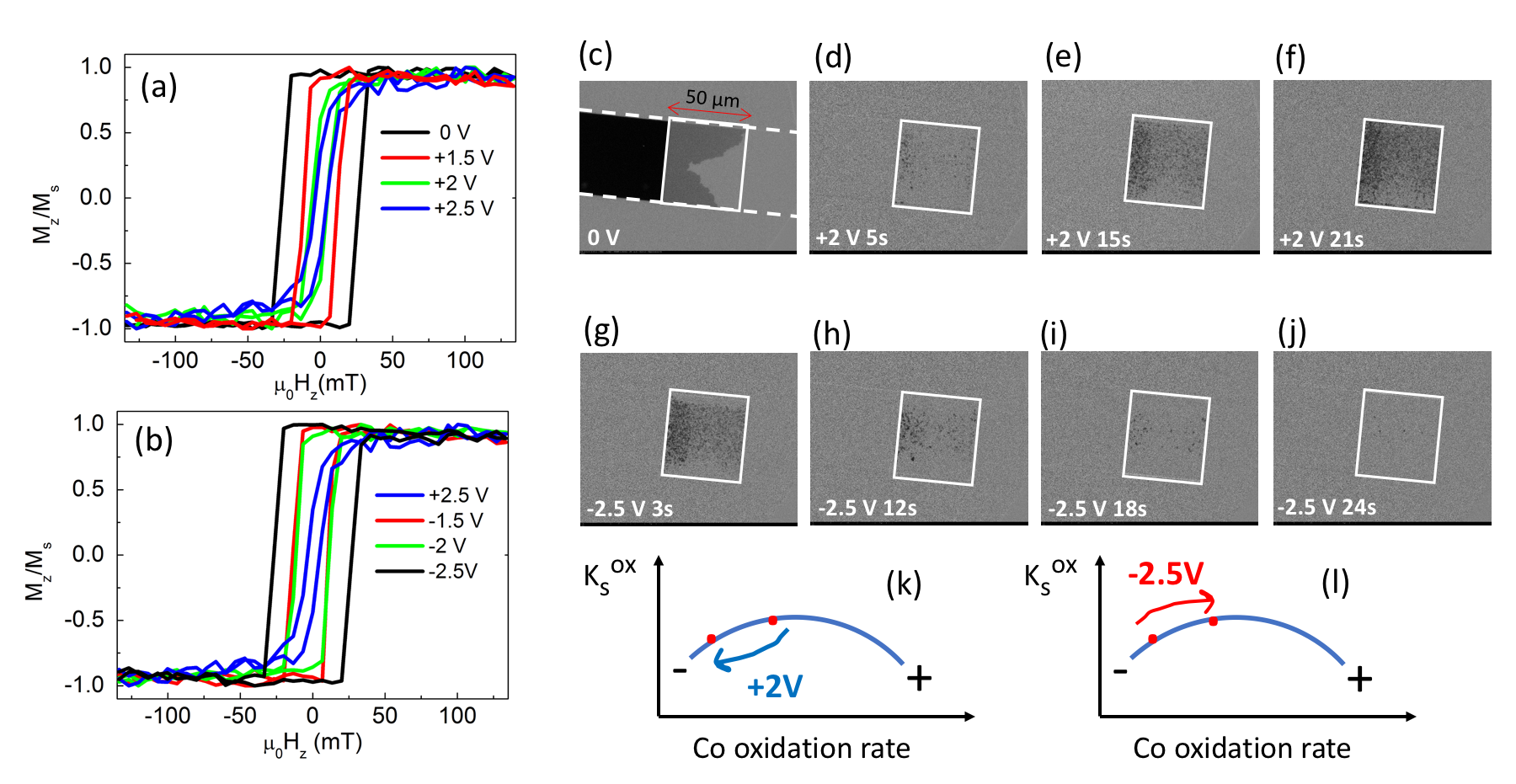}
\end{center}
\caption{\label{fig:Figure2} \textbf{Non-volatile and reversible effect of the electric field in the Pt/Co/AlOx magnetic stack}
(a) hysteresis loops measured below the top Pt electrode in the initial state (before applying the electric field) and after the application of  positive bias voltages $V_{g}$=+1.5~V, +2~V, +2.5~V for 30~s, leading to the decrease of the interfacial PMA; (b)  hysteresis loops measured after the application of  negative bias voltages $V_{g}$=-1.5~V, -2~V, -2.5~V for 30~s, leading to the increase of the anisotropy; the initial state of this series of loops is that obtained after the application of $V_{g}$=+2.5~V for 30~s  in (a);  (c) differential Kerr images showing the reversal by domain wall propagation in the initial state; (d-j) differential Kerr images showing the effect of positive (c-f) and negative (g-j) bias voltage on the nucleation rate of reversed domains. Each hysteresis loop and each image was taken \textit{after} the application of the electric field \textit{i.e.}  with $V_{g}$=0; the nucleation of the reversed  domains (d-j) was by applying one 26 mT field pulse for 20~ms; for (c),  several pulses were applied to bring the domains within the electrode region. The white squares indicate the local electrodes within the magnetic stripes (dotted lines);  (k-l) sketches representing the variation of the interfacial anisotropy constant K$_{s}^{ox}$ \textit{vs}. the Co layer oxidation rate, and the effect of the positive (k) and negative (l) gate voltage.   }
\end{figure*}

\section{Non-volatile and reversible manipulation of interfacial PMA in  P\lowercase{t}/C\lowercase{o}/A\lowercase{l}O\lowercase{x} }
In the initial state, the Co layer in Pt/Co/AlO$_{x}$ presents an out-of-plane magnetisation, with a square hysteresis loop showing a saturated magnetic state at remanence (Fig.~\ref{fig:Figure2}(a)). Following the application of positive voltages ranging from +1.5~V to +2.5~V for 30 seconds, the hysteresis loops (always measured after the voltage was removed, V$_{g}$=0) illustrate a progressive decrease of the coercive field, indicating a decrease of the interfacial PMA. As shown in Fig.~\ref{fig:Figure2}(k), this points to a decrease of the oxidation of the Co layer.  When the reverse voltage is applied (from -1.5 V to -2.5V for 30 seconds, Fig.~\ref{fig:Figure2}(b)) the opposite process occurs:  the gradual widening of the hysteresis loops suggests  the increase of PMA, related to the increase of the Co oxidation (Fig.~\ref{fig:Figure2}(l))  up to the recovery of an hysteresis loop with a coercive field similar to the initial one. The effect is therefore non-volatile and reversible.

\begin{figure*}[ht]
\begin{center}
\includegraphics[width=18cm]{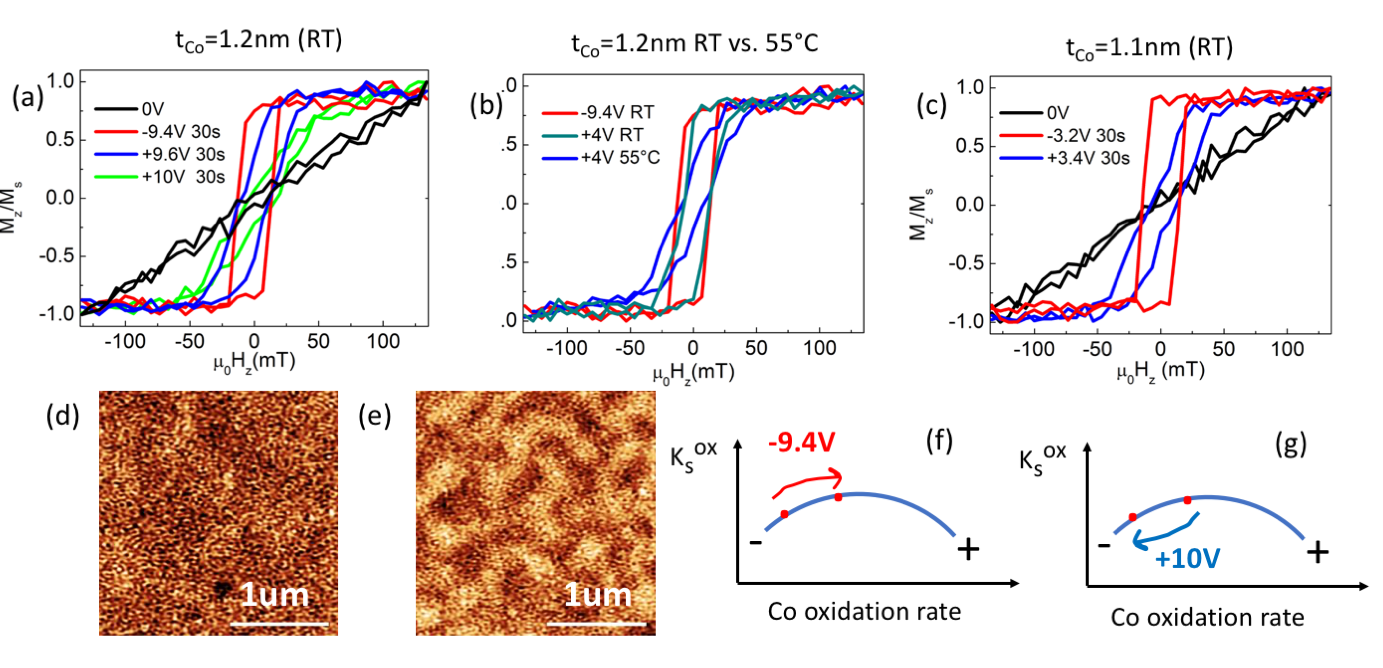}
\end{center}
\caption{\label{fig:Figure3} \textbf{Non-volatile and reversible effect of the electric field in the Pt/Co/TbO$_{x}$ magnetic stack}
(a) hysteresis loops measured in the initial state (IP magnetisation), after the application of  V$_{g}$=-9.4~V for 30~s (increasing PMA) and  V$_{g}$=+9.6~V and +10~V (decreasing PMA) for 30~s  at room  temperature (RT) for a stack with $t_{Co}$=1.2~nm; (b) starting from the state with increased PMA, comparison between the hysteresis loops obtained after the application of V$_{g}$=+4~V at RT and at 55 \degree C for 30~s for the stack with $t_{Co}$=1.2~nm; (c) hysteresis loops measured in the initial state (IP magnetisation), after the application of  V$_{g}$=-3.2~V for 30~s (increasing PMA) and  V$_{g}$=+3.4~V (decreasing PMA) for 30~s  at room  temperature (RT) for a stack with $t_{Co}$=1.1~nm; (d) MFM image obtained after the application of $V_{g}$ = -9.4~V for 30~s leading to saturated OOP state;  (e) MFM image obtained after the application of $V_{g}$ = +10~V for 30~s leading to labyrinthine domains. Dark and light contrasts correspond to down and up magnetisation.  }
\end{figure*}
The  modification of the interfacial PMA can also be deduced from the voltage-induced variation of the  magnetisation reversal process, as illustrated  by the differential  MOKE microscopy images shown in Fig.~\ref{fig:Figure2}(c-j). In the initial state (Fig.~\ref{fig:Figure2}(c)) the magnetisation reversal occurs \textit{via} the expansion of a few large domains, as expected in magnetic layers with sufficiently large anisotropy. The magnetic configurations shown in Fig.~\ref{fig:Figure2}(d-j)  were obtained as follows: after the application of a bias voltage $V_{g}$=$\pm$2~V for a time \textit{t} from 2 to 24~s, the samples were magnetically saturated in the OOP direction with a field +$B_{z}$. A 20~ms long, 26~mT strong  pulse -$B_{z}$ was then applied to reverse the magnetisation. The images show that after the application of a positive voltage $V_{g}$=+2~V,  the magnetisation reversal mechanism changes and is dominated by the nucleation of a large density of domains, the domain density increasing as the application time of the voltage increases. This may be explained within the droplet model \cite{Barbara1978} that predicts that the height of the energy barrier for nucleation is $\Delta E_{n} = \pi\sigma^{2}t/2M_{s}\mu_{0}H $, where $\sigma$ is the domain wall energy, $t$ is the film thickness and $\mu_{0}H$ is the applied magnetic field \cite{Vogel2006}. Since  $\sigma$  decreases when the effective anisotropy decreases, nucleation may dominate the magnetisation reversal in disordered samples with low average magnetic anisotropy, where the  large distribution of anisotropy fields (i.e. of local defects)  may hinder the domain wall movement. As the application time  of the positive voltage increases, the PMA progressively decreases, leading to an increase of the density of the reversed domains \cite{Labrune1989,Pommier1990}. Note that no nucleation event is  observed outside the electrode region, where the PMA remains large and the applied field pulse is not sufficient to overcome the nucleation barriers. When the negative bias voltage is applied for increasing times, starting from the configuration with the lowest PMA (Fig.~\ref{fig:Figure2}(d)),  the density of reverse domains gradually decreases, in agreement with the increase of the PMA suggested by the hysteresis loops.

\section{Non-volatile and reversible manipulation of interfacial PMA in P\lowercase{t}/C\lowercase{o}/T\lowercase{b}O\lowercase{x}}

Fig.~\ref{fig:Figure3}(a) shows the polar MOKE hysteresis loops acquired  for Pt/Co/TbO$_{x}$ in a sample region with $t_{Co}$=1.2~nm. Before applying the electric field, the loop is tilted, as expected for in-plane magnetisation. After  the application of a bias voltage $V_{g}$=-9.4~V  for 30~s, the hysteresis loop becomes square indicating the increase of PMA and the switch to a remanent saturated out-of-plane magnetisation. By applying an opposite voltage  ($V_{g}$=+9.6~V for 30~s) the anisotropy decreases, giving rise to a loop reminiscent of the presence of labyrinthine domains in zero field. When  $V_{g}$=+10~V is applied for 30~s, the anisotropy can be further decreased leading  to in-plane magnetisation orientation as in the initial state.
\par
The voltage-induced modification of the magnetisation state is non volatile and, like for Pt/Co/AlO$_{x}$, the interfacial magnetic anisotropy  can be tuned in a reversible way.
\par
The MFM images shown in Fig.~\ref{fig:Figure3}(d-e) confirm the modification of the magnetic configuration by the electric-field. The image with homogenous contrast observed after the application of  $V_{g}$=-9.4~V  for 30~s (Fig.~\ref{fig:Figure3}(d)) confirms the saturated  magnetic state at remanence  already indicated by the hysteresis loop. After the application of $V_{g}$=+10~V for 30~s (butterfly-like hysteresis loop in Fig.~\ref{fig:Figure3}(a)), the presence of demagnetised state with labyrinthine domains is in agreement with the decrease of the interfacial PMA.

\par
If the sample temperature is increased to 55\degree C during the application of the electric field, the voltage necessary to switch between in-plane and out-of-plane magnetisation  is of the order of -4~V for 30~s, to be compared to -9.4~V for 30~s at room temperature (Fig.~\ref{fig:Figure3}(b)). The kinetics of the oxidation process is therefore enhanced by the temperature.

When the bias voltage is applied to the sample with a thinner Co layer ($t_{Co}$=1.1~nm),~ the voltage necessary to switch the magnetisation from in-plane to out-of-plane at room temperature is reduced to $V_{g}$=-3.4~V for 30~s (Fig.~\ref{fig:Figure3}(c)).  This results can be related to the 1/$t_{Co}$ dependence of the interfacial magnetic anisotropy energy (Eq. 1).

\section{Discussion: microscopic mechanisms}

Our results show that the interfacial magnetic anisotropy at the Co/MO$_{x}$ interface in Pt/Co/MO$_{x}$ trilayers can be strongly  modified by a gate voltage, using a 10nm thick Zr$O_{2}$ dieletric layer deposited by ALD on top of the magnetic stack. A negative/positive gate voltage leads to an increase/decrease of the PMA, which we correlate with an increase/decrease of the oxidation of the initially under-oxidised Co layers.  The efficiency of the effect has been measured to be $>$1200~fJ/Vm for Pt/Co/AlO$_{x}$ at room temperature (see Supplemental Material) and similar values can be expected for Pt/Co/TbO$_{x}$ and Pt/Co/MgO$_{x}$. These values are comparable to the highest reported in the literature
in cases where the gate voltage applied at high temperature triggered oxygen ion migration  \cite{Bi2014,Bauer2015} or where the sample was hydrated giving rise to charge conduction dominated by hydrogen ions \cite{Tan2019a,Tan2019b}. In all these  cases, Gd$_{2}$O$_{3}$   was used as an ion conductor layer.

In Pt/Co/AlO$_{x}$ and Pt/Co/TbO$_{x}$,  the non-volatility and the reversibility of the effect, the large voltage-induced modification of the interfacial PMA at the Co/oxide interface and its dependence on the  strength or the  duration of the applied voltage suggests that  the driving mechanism might  be  the migration of oxygen ions across the ZrO$_{2}$ dielectric layer.

Negative/positive gate voltages, driving O$^{2-}$ ions towards/away from the Co layer, would lead to an increase/decrease of the Co oxidation.  Since the Co layers are initially under-oxidised with respect to the optimal oxidation, that should give rise to an increase/decrease of the interfacial anisotropy, as observed experimentally (see the sketches in Fig.~\ref{fig:Figure1}(b), Fig.~\ref{fig:Figure2}(k,l) and Fig.~\ref{fig:Figure3}(f,g)). These features therefore confirm, as found by previous works \cite{Manchon2008a,Manchon2008b,Manchon2008c,Bernand-Mantel2013} that the modification of the observed PMA is associated with variation of the oxidation of the Co/MO$_{x}$ interface.

The interpretation of our results in terms of oxygen ion migration can also explain the results shown in Figs.~\ref{fig:Figure3}(b-c) for Pt/Co/TbO$_{x}$. When the sample is heated during the application of the electric field, the drift velocity of the oxygen ions is expected to increase \cite{Strukov2009}, so that the same Co oxidation state (i.e. the same change of PMA) can be reached in the same time for a weaker electric field. Similarly, we have observed that the voltage needed to switch the magnetisation from in-plane to out-of-plane decreases as the Co thickness decreases. Because of the 1/t$_{Co}$ variation of the magnetic anisotropy, the needed $\Delta$K$_{s}^{ox}$ i.e. the needed change of Co oxidation, decreases as Co decreases, and can be realised, in the same time, with a lower voltage.

In the case of Pt/Co/MgO$_{x}$ the irreversibility of the electric field effect suggests that the sole oxygen ion migration can not account for the overall process features, and that some irreversible step should occur causing the permanence of the oxidation  of Co at the Co/MgO interface, even after the application of a strong reverse polarisation.

In this work the PMA has been manipulated while maintaining the Co layer under-oxidised. Of course, we expect that the PMA would decrease, as a result of the over-oxidation of the Co layer, if the negative bias voltage was increased or applied for longer times after reaching the maximum anisotropy.

The reversible/irreversible manipulation of the PMA in the systems described in this work could be discussed by considering the Pt/Co/MO$_{x}$/ZrO$_{2}$/Pt (M=Mg,~Al,~Tb) system as a solid state nanometric galvanic cell, where the 10nm thick ZrO$_{2}$ layer  works as a solid electrolyte with pure O$^{2-}$  conductivity and where  the Al,~ Mg or Tb ultrathin oxides work, as well,  as solid electrolytes with pure ionic conductivity or mixed ionic/electronic conductivity.

The oxidation of  Mg, Al and Tb films deposited  above the Co layer is  driven by the large values of the Gibbs free energies of formation of the oxides \cite{Ellingham1944}. For Mg and Al, we expect the formation of stoichiometric (or slightly under-stoichiometric)  MgO and Al$_{2}$O$_{3}$ films while in the case of Tb  the formation of a mixed valence (Tb$^{3+}$ and Tb$^{4+}$) non stoichiometric oxide TbO$_{2-x}$ could be foreseen.
\par
According to results of  Subba Rao \textit{et  al.} \cite{Subba1970}, relative to terbium sesquioxide prepared at high temperature, and to  recent results of Miran et al \cite{Miran2018}, on the presence of surface oxygen vacancies in TbO$_{2}$,  this oxide, like all reduced oxides of the ceria family \cite{Shoko2011}, should present in the bulk phase predominant electronic conductivity, with  significant ionic conductivity associated to oxygen vacancies.
The  presence of more than one phase in nanocrystalline or amorphous terbium oxide \cite{Fursikov2016},  makes it extremely difficult to quantitatively predict its behavior, although  its mixed ionic-electronic conductivity may be safely foreseen, with O$^{2-}$ as the ionic carrier.
\par
The ionic  transport properties of MgO and Al$_{2}$O$_{3}$ are, instead, fully defined, since  nanosized MgO exhibits Mg$^{2+}$ ionic conductivity \cite{Wu2017} and Al$_{2}$O$_{3}$ exhibits  ionic conductivity due to  oxygen ions \cite{Davis1965}, though we could not exclude a small contribution  of electronic conductivity for the non-stoichiometric oxides.

The reversibility associated with the use of Al$_{2}$O$_{3}$ as solid-state electrolyte in the cell Pt/Co/Al$_{2}$O$_{3}$/ZrO$_{2}$/Pt can be, therefore, easily understood if we assume that the top Pt electrode behaves as a sink for atmospheric oxygen, and  works as a reversible oxygen electrode
\begin{equation}
O(Pt)~+2e \leftrightarrow O^{2-}
\end{equation}
at room temperature, with the formation of oxygen ions  O$^{2-}$. These can be transferred, in the case of forward polarisation (negative $V_{g}$) in the ZrO$_{2}$ layer which enables oxygen ion transfer, and then  to the Al$_{2}$O$_{3}$ layer and at the Al$_{2}$O$_{3}$/Co interface,  where each  incoming oxygen ion can form hybrid bonds  with a Co surface atom:
   \begin{equation}
Co~+~ O^{2-} \leftrightarrow O_{Co}^{surf}~+~2e
\end{equation}
where O$_{Co}^{surf}$ is an oxygen bonded  at the Co top surface of the Co electrode.

If the process in Equation 4 is also reversible, under reverse polarisation (positive $V_{g}$)  oxygen can be transferred backwards to the top Pt electrode and the surface of Co is deoxidized, in agreement with the experimental results.

\par
For TbO$_{2-x}$  in the Pt/Co/TbO$_{2-x}$/ZrO$_{2}$/Pt cell we may, again, assume the occurrence of the same reversible electrode reactions and transfer processes of O$^{2-}$ ions, but in parallel with the direct Co oxidation reaction (Eq. 4) a parasitic reaction involving the non stoichiometric Tb oxide may occur, with the TbO$_{2-x}$ layer working as a parasitic oxygen electrode, whose behaviour can be formally described by the following reversible reaction:

\begin{equation}
TbO_{2-x}~+~2ye  \leftrightarrow TbO_{2-z}~+~yO^{2-}
\end{equation}
where z=x-y, that leads to a minute change of the stoichiometry of the Tb oxide.
Therefore, also with the use of TbO$_{x}$ we expect a full reversibility of the entire process, but with a Co oxidation efficiency lower than in the case of the use of Al$_{2}$O$_{3}$ as electrolyte, due to the  parasitic process.
\par
The same overall oxygen transport features can  not be expected to occur in the case of a MgO$_{x}$ layer, due to its Mg$^{2+}$ ionic conductivity \cite{Wu2017}, but the non-volatile and  irreversible oxidation of the Co surface under forward bias may still be understood.
While we expect the reaction in Eq. 3 to occur at the reversible Pt electrode, with the formation of oxygen ions O$^{2-}$, and their further transfer in the ZrO$_{2}$ layer down to the ZrO$_{2}$/MgO$_{x}$ interface, the oxygen transport is blocked in the MgO$_{x}$ layer.

The process occurring at the Co electrode/MgO$_{x}$ interface can be described  with  the following electroneutral equation:
\begin{equation}
Co~+~xMgO  \rightarrow xO_{Co}^{surf}~+~xMg^{2+}~+~2xe
\end{equation}
that induces a surface oxidation of Co, via  the partial electro-reduction of the  MgO$_{x}$ phase.
If we further assume that an ionic carrier interchange occurs at the ZrO$_{2}$/MgO$_{x}$ interface,
where O$^{2-}$ ions do interact with the incoming Mg$^{2+}$ ions arriving from the Co/MgO$_{x}$ interface, with
the intermediate formation of a MgO phase, this would enable an ionic current to flow across the entire cell.

The reaction in Eq. 5  could occur as a reversible process, under severe kinetic hindrances. However, the carrier interchange at the ZrO$_{2}$/MgO$_{x}$ interface, under reverse bias, is forbidden, since the MgO$_{x}$ phase at the ZrO$_{2}$/MgO$_{x}$ interface can not behave as a source of oxygen ions, thus  not enabling an ionic current to flow across the entire cell.
Therefore, we cannot expect that under reverse polarisation the system can restore the primitive configuration, in good agreement with the experimental results.

The specificity of the Pt/Co/MgO$_{x}$/ZrO$_{2}$/Pt system in which we have found an irreversible effect of the electric field, may therefore be related to the different nature of the ionic conduction within the MgO$_{x}$ oxide layer, where it is due by Mg$^{2+}$ cations.

\par
In conclusion, the perpendicular magnetic anisotropy at the Co/MO$_{x}$ interface in Pt/Co/MO$_{x}$ trilayers was modified by an electric field using a ZrO$_{2}$ layer working as a solid electrolyte. Tuning the magnetic anisotropy allowed us to stabilise a variety of magnetic configurations within the Co layers and to modify the details of the magnetic reversal mechanisms. The important efficiency and the non-volatility of the effect is explained in terms of  the migration of oxygen ions towards/away from the Co/MO$_{x}$ interface.  While the effect is reversible in Pt/Co/AlO$_{x}$ and Pt/Co/TbO$_{x}$, where the Co layer can be oxidised or reduced, in Pt/Co/MgO$_{x}$ the effect has been found to be irreversible.  We have attributed these differences to the different nature of the ionic conduction within the MO$_{x}$ layer.  The large voltage control of magnetic anisotropy  ($\beta>$1200fJ/Vm in Pt/Co/AlO$_{x}$) realised with the sample maintained at room temperature, is comparable to that reported in the literature using much higher temperatures \cite{Bi2014,Bauer2015} or in hydrated  samples \cite{Tan2019a,Tan2019b}. This encourages us to continue exploring the use of ZrO$_{2}$ in view of a faster manipulation of interfacial anisotropy.

\section{Acknowledgements }
We acknowledge the support of the Agence Nationale de la Recherche (projects ANR-17-CE24-0025 (TOPSKY), ANR-16-CE24-0018 (ELECSPIN) and ANR-19-CE24-0019 ADMIS) and of the DARPA TEE program through Grant No. MIPR HR0011831554. J.P.G. acknowledges the European Union’s Horizon 2020 research and innovation program under Marie Sklodowska-Curie Grant Agreement No. 754303 and the support from the Laboratoire d’excellence LANEF in Grenoble (ANR-10-LABX-0051). B. Fernandez, Ph. David and E. Mossang are acknowledged for their technical help. We thank Olivier Fruchart for introducing  A.F. to the world of MFM experiments.

\end{document}


\
\section{Sample preparation details}

The magnetic layers were deposited by magnetron sputtering on Si/SiO$_{2}$ 4" wafers at room temperature.  For the Ta(3)/Pt(3)/Co(t)/Mg(0.9) sample (thicknesses in nm), the Co layer was grown as a wedge with thickness \textit{t} varying between 0.77 and 1.5nm.  The Mg layer was oxidised in 150~mbar O$_{2}$ atmosphere for 10~s. A 1.5~nm Ta was deposited on top to prevent
further oxidation : the Ta is expected to oxidise over 1~nm. In the region of the wafer described in this work the Co layer is 1~nm  thick. In the
Pt(3)/Co(0.6)/Al(t) sample the Al layer was grown as a wedge with thickness \textit{t} varying between 0.76 et 2.01~nm. In the region described in this work the Al thickness is 1~nm. The Al layer was oxidised in O$_{2}$ plasma (85 sec, 10~Watt, P~=~3 x 10$^{-3}$ mbar). The Pt/Co/Tb was deposited with both Co and Tb in the form of a
wedge with t$_{Co}$=0.77-1.5~nm and t$_{Tb}$=0.3-0.9~nm. In the region described in this work t$_{Co}$=1.1 to 1.2~nm and t$_{Tb}$=0.6~nm. The Tb was
oxidised in air in the same conditions as the Mg layer. The TbO$_{x}$ layer was covered with a 0.5~nm Al layer that oxidised in air.

\begin{figure}[ht]
\begin{center}
\includegraphics[width=9cm]{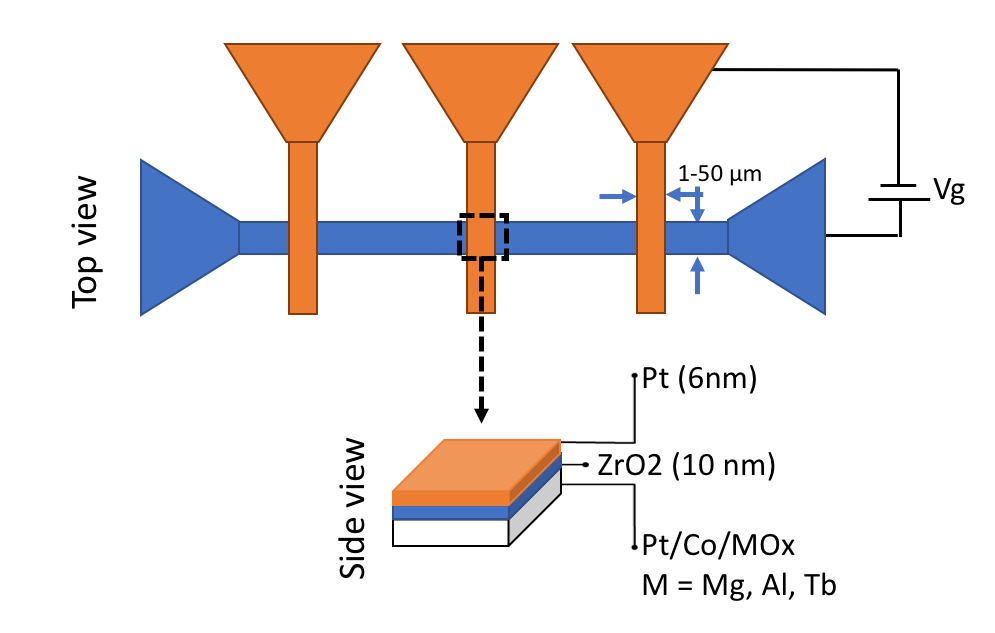}
\end{center}
\caption{\label{fig:FigS1} Sketch of the samples after patterning by electron beam lithography and etching.  }
\end{figure}

The samples were cut into 1cm$\times$1cm pieces and patterned into 1$\mu$m to 50$\mu$m stripes by electron beam lithography.
A 10~nm thick Ti layer was
deposited as hard mask to define the patterned region after lift-off. The samples were then etched in an Ar plasma ; the etching time was calibrated
so that all the Ti layer and part of the capping layers were etched away, in order to terminate close to the top of the MO$_{x}$ layer (M~=~Mg,~Al,~Tb).  We cannot exclude that a small part of the Ta capping layer on top of the MgO
layer has remained after the etching. If a small part of the initial 1.5~nm Ta layer was left, it would be totally oxidised after etching and air
exposure. Nevertheless, we have neglected its presence in the discussion of the main manuscript as Ta$_{2}$O$_{5}$ behaves as   an oxygen ion conductor similarly to
the ZrO$_{2}$ layer. If the Ta was not oxidised and behaved as a barrier for oxygen ion migration, the large electric field effect on PMA observed for
negative voltage could not be explained.
 A 10~nm thick
ZrO$_{2}$ layer was then  deposited on top of the whole sample, using atomic layer deposition (ALD) (0.1~nm/cycle at 100 \degree C using TDMAZr/H$_{2}$O precursors). The sample temperature during the soft baking process was kept at  100\degree C to avoid
degradation of the  magnetic properties induced by interface mixing. X-ray diffraction shows that the layer is amorphous.
 After the deposition of the ZrO$_{2}$ layer, a second lithography step was carried out to define the top
electrodes,   obtained by lift-off after evaporation of a 6~nm Pt thick layer.
The electric field was applied keeping the top Pt electrically grounded. This set-up was chosen to avoid an electrostatic interaction between the sample
and the MFM tips. A sketch of the sample after patterning in shown in Fig.~\ref{fig:FigS1}.

In some of the experiments, the gate voltage was applied with the sample being heated, while keeping it in an ambient atmosphere.

\section{Micromagnetic simulations}

\begin{figure}[ht]
\begin{center}
\includegraphics[width=12cm]{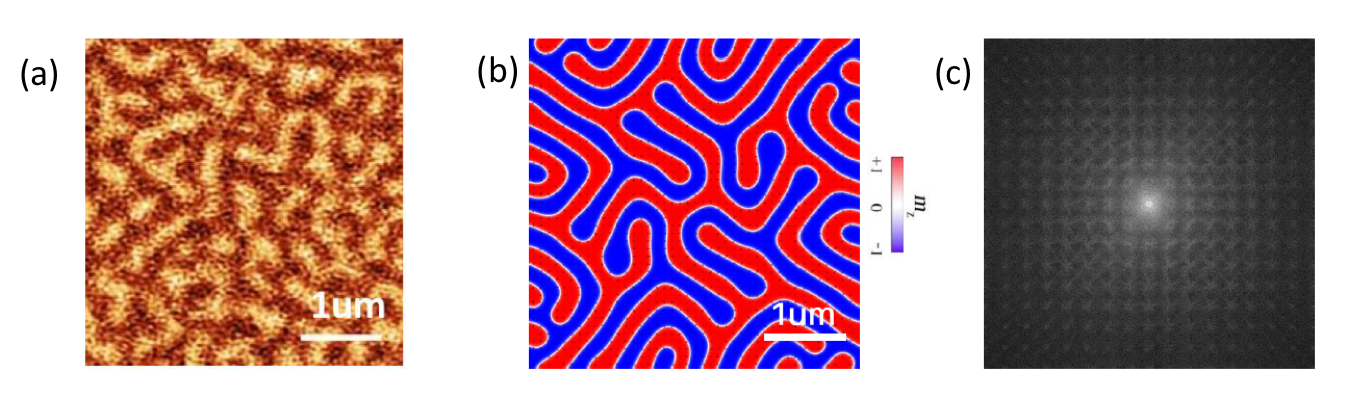}
\end{center}
\caption{\label{fig:FigS2} (a) Measured stripe domains observed for the Pt/Co/MgO sample under the 50$\mu$m$\times$5$\mu$m electrode by MFM. The strip domains show a periodicity of.... nm; (b) Simulated stripe domains with a periodicity of .... nm obtained with the same magnetic paramters of a). . (c) Fourier transform of b)}
\end{figure}

The stripe domain configuration observed by MFM before the application of gate voltage (Figure \ref{fig:FigS2}(a)  is  determined  by the minimization  of  the  total  energy  including  demagnetizing and  domain  wall  energies \cite{Moreau2016}. Micromagnetic simulations were performed using the  MuMax3 software package \cite{Vansteenkiste2014,Mulkers2017} in order to reproduce this magnetic domain configuration.

The magnetic parameters were chosen from  previously published experimental data for a sample prepared in the same conditions \cite{Juge2019}.  We used: $A$ = 16 pJ/m, $M_{s}$ = 1.42~MA/m, $K_{u}$ = 1.34 MJ/m$^{3}$. The DMI strength $D$ was varied between 1.2~mJ/m$^{2}$ and  1.4~mJ/m$^{2}$ in steps of 0.05~mJ/m$^{2}$. The best agreement with the experimental results, was found with $D$=1.3~mJ/m$^{2}$, a value similar to that reported for this system \cite{Juge2019}. The simulated sample was discretised in rectangular cells of $(1,1,\mathrm{t_{Co}})\textrm{nm}^3$ where $\mathrm{t_{Co}}$ is the thickness of the cobalt layer in the experiment (1nm). The  lateral  dimensions  of  the  simulation  cell  were  chosen  to be  a  few  times smaller  than  the anisotropy exchange length $\Delta =\sqrt{A/K_{eff}}= 14.8~\textrm{nm}$ and the dipolar exchange length $ \delta_{D} =\sqrt{2A/\mu_0 M_s^2}= 3.6 \textrm{nm}$. We considered a region of 4096 $\textrm{nm}^3$ (gridsize of (4096,4096,1)), with periodic boundary conditions along the x and y directions in order to reproduce an infinite thin film. Finally, the damping parameter was set to $\alpha=0.5$ in order to speed up convergence.

The initial magnetic configuration was a 200~nm diameter bubble domain. The magnetization was then relaxed at zero external field  until the energy has reached a stationary state and the maxtorque is lower than $10^{-5}$ Tesla. The equilibrium labyrinth domain width  was extracted by  taking  a  Fourier transform. The Fourier transform intensity was radially averaged and the peak was fitted to a Gaussian function.

Fig. \ref{fig:FigS2}(b) shows the relaxed stripe domain state for $D=1.3 \mathrm{mJ/m^2}$, with a equilibrium domain width of 115~nm jointly with his Fourier transform (Fig. \ref{fig:FigS2}(c)).  This domain width is in good agreement with the experimental value of 120~nm.

\section{Determination of the efficiency of the voltage control of magnetic anisotropy (VCMA) effect}

\begin{figure}[ht]
\begin{center}
\includegraphics[width=15cm]{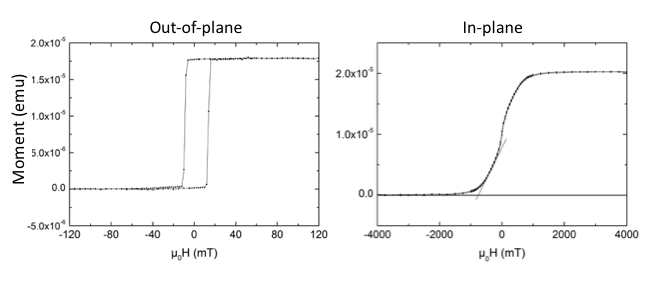}
\end{center}
\caption{\label{fig:FigS3} Easy and hard axis hysteresis loops measured for Pt/Co/AlO$_{x}$}
\end{figure}

The efficiency of the VCMA effect (called $\beta$ factor in the literature) is defined as the variation of the interfacial magnetic anisotropy
constant $K_{s}$  per unity of electric field (1~V/m).
Values of the order of several tens or hundreds fJ/Vm have been reported in the literature when the effect is due to pure charge accumulation/depletion, while
values of several 1000~fJ/Vm have been reported when the oxidation of the Co/oxide interface is driven by oxygen ion migration.
In order to evaluate the $\beta$ factor for the Pt/Co/AlO$_{x}$, the easy-axis and hard axis hysteresis loops of the magnetic layer was  measured  by
VSM-SQUID in a region of the 4 inch wafer similar to the one of the patterned sample. The measurements presented in Fig.~\ref{fig:FigS3} allowed to
extract the spontaneous magnetisation of the sample $M_{s}$=1.26~MA/m  and in-plane saturation field $\mu_{0}H_{k}$=0.76~T, from which we extracted
the effective magnetic anisotropy $K_{eff}$=0.48 MJ/m$^{3}$. From the expression:
\begin{equation}
 K_{eff} = \frac{K_{s}}{t_{Co}} -\frac{1}{2}\mu_{0} M_{s}^{2}
 \end{equation}
and t$_{Co}$=0.6nm, we extract $K_{s}$=0.88 10$^{-3}$ J/m$^{3}$.

The application of an electric field of (2.5V/10nm)  allowed us to decrease the anisotropy down to that of close to the spin reorientation transition
(where the sample demagnetises in stripe domains) \textit{i.e.} where  $K_{eff}$=0 and therefore where $K_{s}$=0.59 $\times$ 10$^{-3}$ J/m$^{3}$. This
leads to $\beta~>~$1160 fJ/Vm.

